\documentclass[pdflatex,sn-mathphys-num]{sn-jnl}

\usepackage{graphicx}%
\usepackage{multirow}%
\usepackage{amsmath,amssymb,amsfonts}%
\usepackage{amsthm}%
\usepackage{mathrsfs}%
\usepackage[title]{appendix}%
\usepackage{xcolor}%
\usepackage{textcomp}%
\usepackage{manyfoot}%
\usepackage{booktabs}%
\usepackage{algorithm}%
\usepackage{algorithmicx}%
\usepackage{algpseudocode}%
\usepackage{listings}%
\usepackage{braket} 
\usepackage[separate-uncertainty=true]{siunitx}

\theoremstyle{thmstyleone}%
\theoremstyle{thmstyletwo}%

\theoremstyle{thmstylethree}%

\begin{document}

\title[Article Title]{Finite-momentum Cooper plasmons in superconducting terahertz microcavities }

\author[1,2,3,4]{\fnm{Alex M.} \sur{Potts}}
\equalcont{These authors contributed equally to this work.}

\author[1,2,4]{\fnm{Marios H.} \sur{Michael}}
\equalcont{These authors contributed equally to this work.}

\author[1,2]{\fnm{Gunda} \sur{Kipp}}
\author[1,2]{\fnm{Sara M.} \sur{Langner}}
\author[1,2]{\fnm{Hope M.} \sur{Bretscher}}
\author[3,6]{\fnm{Jonathan} \sur{Stensberg}}
\author[1,2]{\fnm{Kelson} \sur{Kaj}}
\author[1,2]{\fnm{Toru} \sur{Matsuyama}}
\author[1,2]{\fnm{Matthew W.} \sur{Day}}
\author[1,2,3]{\fnm{Felix} \sur{Sturm}}
\author[4]{\fnm{Abhay K.} \sur{Nayak}}
\author[4]{\fnm{Liam A.} \sur{Cohen}}
\author[3,6]{\fnm{Xiaoyang} \sur{Zhu}}
\author[4]{\fnm{Andrea} \sur{Young}}
\author*[1,2,3]{\fnm{James} \sur{McIver}}\email{jm5382@columbia.edu }

\affil[1]{\orgname{Max Planck Institute for the Structure and Dynamics of Matter}, \orgaddress{\street{Luruper Ch 149}, \city{Hamburg}, \postcode{22761}, \state{Hamburg}, \country{Germany}}}

\affil[2]{\orgname{Center for Free-Electron Laser Science (CFEL)}, \orgaddress{\street{Luruper Ch 149}, \city{Hamburg}, \postcode{22761}, \state{Hamburg}, \country{Germany}}}

\affil[3]{\orgdiv{Department of Physics}, \orgname{Columbia University}, \orgaddress{ \street{538 West 120th Street}, \city{New York City}, \postcode{10027}, \state{New York}, \country{United States}}}

\affil[4]{\orgdiv{Department of Physics}, \orgname{University of California Santa Barbara}, \orgaddress{ \city{Santa Barbara}, \postcode{93106}, \state{California}, \country{United States of America}}}

\affil[5]{\orgname{Max Planck Institute for the Physics of Complex Systems}, \orgaddress{ \street{Nöthnitzer Straße 38}, \city{Dresden}, \postcode{01187}, \country{Germany}}}

\affil[6]{\orgdiv{Department of Chemistry}, \orgname{Columbia University}, \orgaddress{ \street{3000 Broadway}, \city{New York City}, \postcode{10027}, \state{New York}, \country{United States}}}


\abstract{The phase mode of a superconductor's order parameter encodes fundamental information about pairing and dissipation, but is typically inaccessible at low frequencies due to the Anderson-Higgs mechanism. Superconducting samples thinner than the London penetration depth, however, support a gapless phase mode whose dispersion can be reshaped by a proximal screening layer. Here, we theoretically and experimentally show that this screened phase mode in a superconducting thin film integrated into on-chip terahertz circuitry naturally forms a superconducting microcavity that hosts resonant finite-momentum standing waves of supercurrent density, which we term Cooper plasmons. We measure two Cooper plasmons in a superconducting NbN microcavity and demonstrate that their resonance frequencies and linewidths independently report the density of participating carriers and plasmon's dissipation at finite momenta. Our results reveal an emergent collective mode of an integrated superconductor–circuit system and establish design principles for engineering or suppressing Cooper plasmons in superconducting terahertz devices and circuits.

}


\maketitle
 
\section{Introduction}\label{sec1}

\begin{figure}[h]
    \centering
    \includegraphics[width=0.5\textwidth]{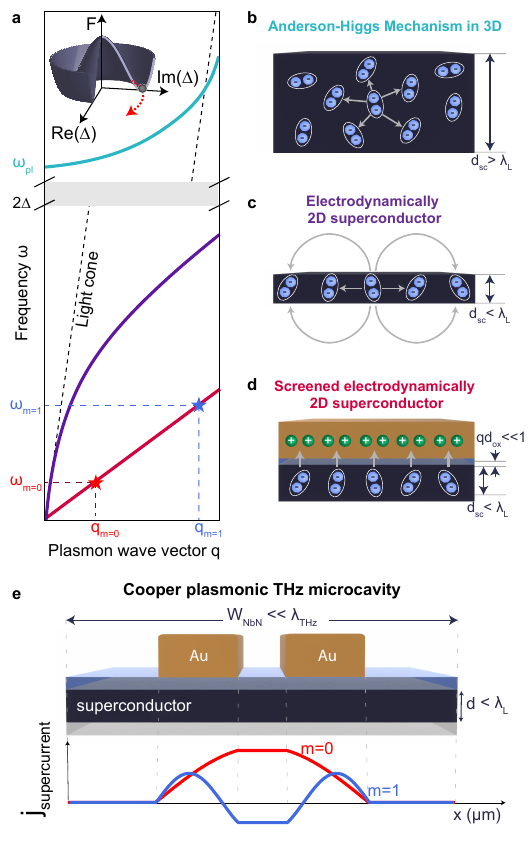}
    \caption{ \textbf{Summary of dispersion relations} (a) Dispersion relations for the phase mode of the pictured superconducting samples, which take different functional forms due to dimensionality and screening. The dispersions of both the electrodynamically two-dimensional and screened electrodynamically two-dimensional samples are below the light cone. (inset) Landau-Ginzburg potential, with amplitude and phase modes indicated. (b) Electrodynamically three-dimensional superconducting sample, where Cooper pairs interact isotropically with other Cooper pairs (c) Electrodynamically two-dimensional superconductor with both local in-plane and non-local (long-range) out-of-plane interactions. (d) Electrodynamically two-dimensional superconductor with proximal metal screening layer, separated by an oxide.  (e) Schematic of a superconducting microcavity capacitively contacted to a gold coplanar stripline (CPS). The superconductor is thinner than the London penetration depth and narrower ($W_{NbN}$) than the wavelength of THz photons. Pictured are two Cooper plasmons (red: lower frequency mode and blue: higher frequency mode), which are resonant current distributions formed of Cooper pairs. }\label{Fig1_phaseMode}
\end{figure}

Superconductivity is one of the most famous macroscopic quantum phases in many-body systems. The thermodynamic properties of most superconductors are well-described by the Landau-Ginzburg free energy, which exhibits a ``Mexican hat" potential in the real and imaginary parts of the superconducting order parameter (Fig.~\ref{Fig1_phaseMode}a and inset). This potential supports two low-energy collective modes \cite{shimano2020}: a radial (amplitude / Higgs) mode and an azimuthal (phase / Nambu-Goldstone) mode. The Higgs mode has an energy scale of twice the superconducting gap (2$\Delta$) and has been accessed by nonlinear far-field optical probes; the phase mode is the long-wavelength Goldstone mode associated with spontaneous symmetry breaking. In conventional three-dimensional (3D) superconductors, Coulomb interactions between Cooper pairs (Fig. \ref{Fig1_phaseMode}b) force the phase mode to acquire an eV-scale energy gap by the Anderson-Higgs mechanism \cite{anderson1963} and are responsible for the Mei${\ss}$ner effect, wherein incident magnetic fields are exponentially attenuated with a length scale of the London depth $\lambda_{\text{L}}$. This energy gap is much larger than typical values of the superconducting gap (2$\Delta$ $\sim$ a few meV), meaning that eV-scale photons will destroy the superconducting order before the phase mode dynamics can be probed, making the phase mode seldom studied experimentally. 

The nature of Coulomb screening, however, changes when the superconductor thickness ($d_{sc}$) is lowered below $\lambda_{L}$ which, in turn, changes the dispersion of the phase mode (see Appendix A). Local Coulomb screening in three dimensions is replaced by both local in-plane and nonlocal out-of-plane screening in two dimensions (Fig.~\ref{Fig1_phaseMode}c) and, thus, samples transition from being electrodynamically three-dimensional (3D) to two-dimensional (2D). The exponential decay of incident fields with the London depth length scale is replaced by a power law decay set by the Pearl length ($\lambda_{p} = 2\lambda_{L}^2/d_{sc}$), which is often several microns \cite{pearl1964,irz1995,ilin2014}. This change to the Coulomb kernel suppresses the conventional Anderson-Higgs mechanism, allowing the phase mode to remain gapless \cite{Sun2020,diamantini2022,li2025}. The gapless phase mode is infrared active in the far-field and has a dispersion relation given by $\omega_{p} \sim \sqrt{ q/\lambda_{p}}$, with frequency $\omega_{p}$ and in-plane momentum $q$ \cite{Sun2020}. This dispersion is further modified by a metal screening layer placed at distance $d_{ox}$, converting the $\sqrt{q/\lambda_p}$ dispersion to one that is linear in momentum ($\omega_p \sim q\sqrt{d_{ox}/\lambda_p}$). The screened and unscreened phase modes of electrodynamically 2D superconductor can, therefore, be probed at energies less than 2$\Delta$, making them experimentally accessible. This dispersion, however, lies beyond the light cone such that a technique with finite-momentum transfer is required.

Near-field techniques couple light to finite-$q$ excitations via spatial confinement and allow for spectroscopically probing superconductors beyond the light cone \cite{Stinson2014, Sun2020, Berkowitz2021, Costa2021, li2025, schumacher1998, dunmore1995, buisson1994, hoegen2025}. One such technique is on-chip THz spectroscopy, which leverages transmission line circuitry to interface THz electric fields with subwavelength samples \cite{Wood2013, Gallagher2019, Yoshioka2022, Zhao2023, Yoshioka2024, kipp2025,kusyak2025}, including superconductors \cite{Potts2023a, Seo2024, Wang2023, adelinia2025}. On-chip THz spectroscopy has recently revealed self-cavity effects \cite{kipp2025, michael2025, xli2025}, wherein finite momentum standing waves of current density form due to reflection from physical and electrostatic edges. These self-cavity momenta can be hundreds of times that of a bare photon, allowing for the exploration of collective modes well outside the light cone.   

Here, we theoretically predict and experimentally identify the superconducting phase mode via the on-chip THz spectroscopic response of an electrodynamically 2D NbN self-cavity. The spectrometer's transmission line both enables near-field spectroscopy and electrostatically screens the superconductor, inducing self-cavity effects that imbue the Cooper pairs with finite momentum (Fig.~\ref{Fig1_phaseMode}e). The result is resonant finite-momentum standing waves of supercurrent density at frequencies of a few hundred gigahertz, which we term `Cooper plasmons'. These Cooper plasmons are irreducibly a product of superconductivity, dimensionality, screening, the self-cavity, and circuit design, and are only observable because the integrated device has emergent properties beyond those of any single constituent. Furthermore, these modes are neither exotic nor material-specific: they are a generic feature whenever a superconductor is thinner than a few times the $\lambda_L$ (see Appendix B) and especially when screened by a proximal metal layer, which is common to many superconducting circuits. Cooper plasmons could emerge in any integrated superconducting system with a transmission line \cite{blais2021}, making them a novel and relevant loss channel in superconducting circuits as operating frequencies continue to increase. 

Cooper plasmons can also be used to unlock key information about a superconductor's self-energy. The Cooper plasmon resonance frequency and scattering rate are directly related to the carrier / supercarrier densities and lifetime, which stem from the real and the imaginary parts of the self-energy $\Sigma$. These key quantities describe the renormalization of a single particle's properties by interaction with its quantum environment and are notoriously difficult to measure independently \cite{damascelli2003, Basov2011}. We show that Cooper plasmons effectively sample $\Re[\Sigma]$ and $\Im[\Sigma]$ at the frequency and momentum selected by the microcavity boundary conditions, and we exploit this characteristic to track the opening of the superconducting gap in a thermal superconductor-to-metal transition as a function of both frequency and finite momentum. This ability to distinguish between $\Re[\Sigma]$ and $\Im[\Sigma]$ is a general feature of Cooper plasmons and suggests that they could be powerful reporters of other superconducting collective modes and the systems' underlying Hamiltonians themselves.

\section{Results}\label{sec2}  

\subsection{Cooper plasmons as the superconducting phase mode in a microcavity}\label{subsec2.1}

We theoretically model the response of an ideal superconducting microcavity by describing a superconducting sample of thickness $d_{\text{sc}}$ with rectangular boundary conditions in-plane on a low-loss substrate. An insulating oxide layer of thickness $d_{\text{ox}}$ and relative permittivity $\epsilon_{1}$ is placed above the sample. A coplanar stripline (CPS) transmission line, which is defined by two conductors of width $W$ and separation $S$, is used to transport THz radiation to and from the sample. The superconductor under the metal traces is electrostatically screened; the superconductor not under the metal traces is unscreened. The THz electric field propagating in the CPS has a quasi-transverse electromagnetic (TEM) mode, with its field predominantly confined in the $x$-direction between the metal traces. The quasi-TEM mode injects current into the center unscreened region, which propagates with unscreened momentum $q_u$. This current partially transmits into the screened region, where it carries momentum $q_s$. Current in the screened region feeds back into the unscreened region and the CPS's quasi-TEM mode outcouples radiation, acting as a near-field detector \cite{michael2025}. The resulting microcavity conductivity ($\sigma_{\text{mc}}$) can be written in terms of the intrinsic material conductivity ($\sigma_{\text{in}}$) and a feedback factor $F(\omega, q_u, q_{\text{s}})$, which quantifies self-cavity effects, as: 

\begin{equation}
    \sigma_{\text{mc}}(\omega) = \sigma_{\text{in}}(\omega) \left( 1+ F(\omega, q_u, q_{\text{s}})  \right).
\end{equation}\label{eqn1}

The analytic form of $F(\omega, q_u, q_{\text{s}})$ can be derived for any generic microcavity geometry \cite{michael2025, kipp2025, xli2025, Kipp2025b}, including effects from ohmic charge transfer between the superconductor and CPS via tunneling across the oxide by an oxide conductivity $\sigma_{ox}$. The expression for $F(\omega, q_u, q_{\text{s}})$ for the geometry shown in Fig. \ref{Fig1_phaseMode}d is provided in Appendix C. $F(\omega, q_u, q_{\text{s}})$ goes to -1 when the injected and feedback currents destructively interfere, causing $\sigma_{\text{mc}}(\omega)$ to vanish at certain frequencies in spite of non-zero $\sigma_{\text{in}}(\omega)$. $F(\omega, q_u, q_{\text{s}})$ also has poles, corresponding to resonances in $\sigma_{mc}(\omega)$ with the form of a Lorentzian. These poles physically manifest as plasmons: resonant, standing waves of current density. At zero temperature, the plasmons are wholly comprised of Cooper pairs, as pictured for $m = 0$ (red) and $m=1$ (blue) in Fig.~\ref{Fig1_phaseMode}e. In our geometry, the plasmonic standing wave is dominated by $q_{\text{s}} \sim 50^{-1} \mu m^{-1}$ and oscillates at frequencies on the order of \SI{0.5} THz. This momentum is \SIrange{10}{100}{} times that of a bare photon in vacuum.

\begin{figure}[h]
\centering
    \includegraphics[width=0.5\textwidth]{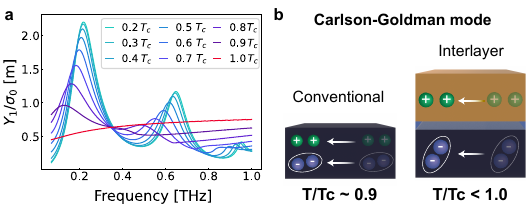}
    \caption{ \textbf{Normalized microcavity conductivity and collective mode structure below $T_c$} (a) Y$_1/\sigma_0$, proportional to the imaginary part of the microcavity conductivity. For comparison to the experiment presented later, we use a mixture of ohmic and capacitive contact ($\sigma_{ox}/\sigma_{in} \approx 0.4$). (b) Comparison of collective modes. A conventional Carlson-Goldman mode peaks at $T/T_c \sim 0.9$, whereas the amplitude of our interlayer Carlson-Goldman (Cooper) plasmons grows as the temperature drops to zero. }\label{Fig2_cgMode}
\end{figure}

We computed the temperature-dependent superconducting $\sigma_{mc}(\omega)$ using an extended Zimmermann model \cite{Zimmermann1991} for $\sigma_{in}(\omega)$, which parametrizes an s-wave superconductor with superconducting gap $2\Delta(T)$, normal metal carrier scattering rate ($\gamma_{\text{nm}}$), and Cooper plasmon relaxation rate ($\gamma_{\text{sc}}$) coming from ohmic leakage to the transmission line or by sidewall loss \cite{kim2025} (see Appendix D). The real part of $\sigma_{mc}(\omega)$ is depicted in Fig. \ref{Fig2_cgMode}a  using $\gamma_{\text{sc}}$ = 145 GHz and $\gamma_{\text{nm}}$ = 5.5 THz \cite{Sim2017}. The plasmon is entirely composed of overdamped normal metal carriers above the critical temperature $T_c$, so plasmonic features are not clearly visible. As the temperature decreases below $T_c$, both highly-damped normal metal carriers and minimally-damped Cooper pairs contribute to a visible Lorentzian lineshape. As the temperature decreases, the plasmonic standing wave becomes increasingly composed of minimally-damped Cooper pairs, which leads to a slight blue-shift and large increase in quality factor \cite{xli2025}. On resonance at zero temperature, the plasmonic standing wave is purely composed of Cooper pairs - forming a Cooper plasmon.

The Cooper plasmon studied here  consists of two types of oscillators coupled by the flow of (super)current. One of the oscillators is the non-resonant $\sqrt{q}$ phase mode in the unscreened region; the other is the resonant linear-in-$q$ phase mode in the screened regions. The resonant contribution from the screened regions originates from an acoustic mode that consists of an antisymmetric oscillation of Cooper pairs in the superconducting layer and image charges in the metal transmission lines. This is analogous to the conventional Carlson-Goldman mode, which is a superconducting collective mode resulting from the movement of Cooper pairs and an ambipolar counterflow of normal metal carriers \cite{carlson1975, schmid1975}. The resonant screened phase mode acts as an interlayer Carlson-Goldman mode (Fig. \ref{Fig2_cgMode}b), with the Cooper pairs and image charge screening normal metal carriers physically separated by $2d_{ox}$. The conventional Carlson-Goldman mode is fragile and difficult to experimentally access \cite{artemenko1979, Sun2020} because it requires a comparable density of both Cooper pairs and normal metal carriers, a condition only fulfilled in S-wave superconductors at temperatures just below T$_c$. In addition, the high scattering rate of the normal metal carriers in dirty superconductors \cite{goldman1981} and aggressive Landau damping \cite{artemenko1979} make the Carlson-Goldman mode overdamped. These factors, combined with the mode's near charge-neutral nature, make the conventional Carlson-Goldman mode optically silent in the far-field and only a weak, overdamped feature in the near-field reflection coefficient \cite{Sun2020}. 

The physical separation between supercarriers and normal metal carriers, combined with the ample supply of normal metal carriers in the screening layer, allows the Cooper plasmon's oscillator strength to grow monotonically as the temperature is lowered. Remarkably, this physical separation also increases the interlayer Carlson-Goldman mode's phase velocity far beyond the normal metal carrier Fermi velocity, such that the Landau damping is minimized. The physical separation of carrier species also makes the Cooper plasmon damping independent of the normal carrier scattering rate, $\gamma_{nm}$, in the metal transmission lines (see Appendix E). As a result, the interlayer Carlson-Goldman mode's robustness makes it a unique platform to probe the superconducting order at finite momentum.

\subsection{On-chip THz spectroscopy}\label{subsec2.2} 

\begin{figure}[h]
\centering
    \includegraphics[width=0.5\textwidth]{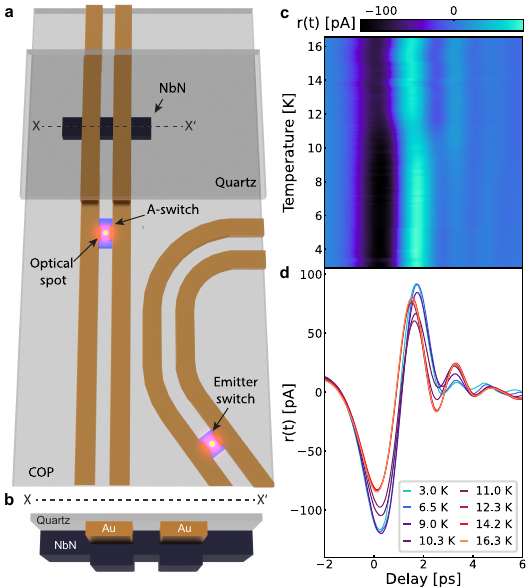}
    \caption{ \textbf{On-chip THz spectroscopy of NbN with the fast sample interchange architecture.} (a) Schematic showing the A switch, the sample board - switch board interface, and the NbN shunt from which the THz transient reflects. The sample board is quartz, the switch board is cyclo-olefin polymer (COP). (b) Cross-section of the sample board along A-A'. (c) Time-domain reflected signal versus temperature with a $T_C$ of 11.1 K, with (d) linecuts at selected temperatures. }\label{Fig3_data}
\end{figure}

To experimentally investigate the Cooper plasmons, we used an on-chip THz spectrometer with a fast-interchange architecture \cite{Lee2022, Potts2023a} and a NbN microcavity (Fig. \ref{Fig3_data}a-b), a dirty-limit Bardeen–Cooper–Schrieffer (BCS) superconductor. A \SI{151}{nm} thick, \SI{8}{\micro\meter} long superconducting NbN sample with \SI{11.1}{K} $T_c$ was fabricated in direct contact with a CPS ($W =$~\SI{50}{\micro\meter}, $S =$~\SI{30}{\micro\meter}) on an interchangeable quartz sample board. The intrinsic conductivity of NbN is well-captured by the Zimmermann model \cite{Zimmermann1991, Sim2017, cheng2016}, in accordance with our model. The NbN's rectangular boundary conditions minimize inhomogeneous broadening and ensure microcavity behavior \cite{Kipp2025b}. The NbN microcavity is thinner than the $\lambda_L$ of \SI{500}{\nano\meter} \cite{Sim2017}, making it electrodynamically 2D and preventing the Anderson-Higgs mechanism from gapping the phase mode. The microcavity selects momenta on the order of 50$^{-1}$ $\mu m^{-1}$, allowing us to measure outside the light cone.

THz radiation is generated by a low-temperature GaAs emitter Auston switch. An Auston switch detector (`A'-switch) then records the THz signal both prior to entering the sample board and following reflection from the device on the sample board. We extracted the directivity and reflection signals in the normal metal (D$_{nm}(\omega)$ and R$_{nm}(\omega)$) and superconducting states (D$_{\text{sc}}(\omega)$ and R$_{\text{sc}}(\omega)$). The time-domain reflected signal is plotted as both a colormap (Fig. \ref{Fig3_data}c) and linecuts at selected temperatures (Fig. \ref{Fig3_data}d). At the $T_c$, the \SI{0}{ps} and \SI{2}{ps} signal components unmistakably shift to later times. In addition, the period and amplitude of the oscillations occurring at and after \SI{3}{ps} change, indicating that the superconducting transition significantly modifies the reflection's spectral weight distribution.

\subsection{Experimental analysis of Cooper plasmons }\label{subsec2.3}  

\begin{figure}[h]
\centering
\includegraphics[width=0.5\textwidth]{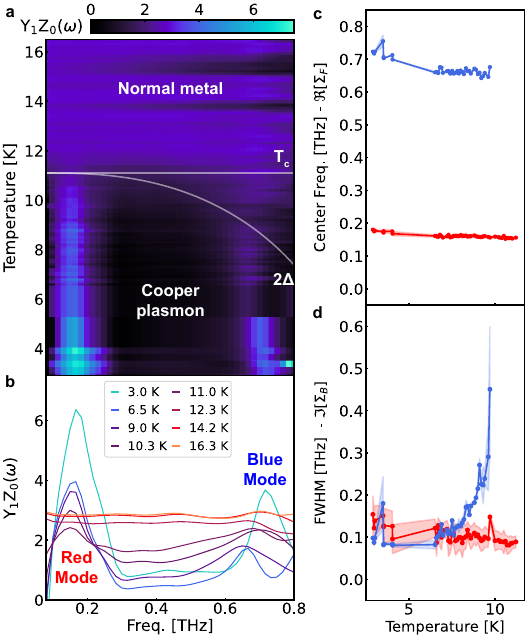}
    \caption{ \textbf{Extracted parameters from fitting experimental data} (a) Calculated optical conductance $Y_1Z_0$ with (b) select linecuts. White lines demarcate $T_C$ and the temperature-dependent superconducting gap $2\Delta(T)$. Strong absorptions develop at 150 GHz (m = 0, red mode) and 620 GHz (m = 1, red mode) upon entering the superconducting condensate. The resulting fitted (c) center frequency and (d) full width, half maximum (FWHM). Shaded regions indicate 1-standard deviation from curve-fitting. Fits to the blue mode above \SI{9.9}{K} were terminated, as the Lorentzians there become too broad to fit properly. }\label{Fig4_fits}
\end{figure}

The normalized sample admittance $Y_{sc}$, which is $\sigma_{mc}$ times geometric factors, can be extracted via eqn. \ref{eqn2} (methods and Appendix F). The normalized conductance Y$_1$Z$_0$, which is proportional to the real part of $Y_{sc}$, exhibits two resonances below $T_c$, which we attribute to finite-momentum Cooper plasmons (Fig. \ref{Fig4_fits}a and b). Only the conductivity of the normal state manifests above $T_c$. As the temperature is lowered below $T_c$, two clear resonances form at approximately \SI{170}{GHz} (m = 0, red mode) and \SI{650}{GHz} (m = 1, blue mode). The amplitude of the resonant peaks increases as $T/T_c$ is lowered, and their resonance frequencies are only weakly temperature-dependent. The conductivity of similarly thick NbN is well-known through far-field measurements \cite{beck2011, pracht2013, cheng2016, Sim2017} and, in contrast, displays no sub-gap collective modes of this magnitude. 

The resonance frequencies do, however, match those of the finite-momentum Cooper plasmons predicted by our theory. We quantitatively examine them by fitting the real part of the admittance on and near resonance to Lorentzian functions, effectively removing both ohmic contact effects (Appendices G and H) and temperature-dependent Zimmermann backgrounds. We extract the center frequency and full width, half maximum (FWHM) of both modes as a function of temperature, which are plotted in Fig.~\ref{Fig4_fits}c and d (see Appendix I). 

The temperature evolution of the Cooper plasmons reflects the intricacies of the superconductor's phase diagram. Above $T_c$, the plasmons are expected to be charge-carrier plasmons, but are overdamped by the large normal metal scattering rate and otherwise suppressed by the ohmic contact. Below $T_c$ and $2\Delta(T)$, $\sigma_{\text{ox}}/\sigma_{\text{in}}$ drops, permitting feedback and allowing plasmons to form. The plasmons become underdamped and composed of a mixture of Cooper pairs and normal metal carriers. As the sample is cooled and normal metal carriers freeze out, the lossiness decreases and the FWHM decreases. The resonance frequencies of both the red and blue mode weakly blueshift by $\sim$15\%, likely due to the slight shift of a resonance's center frequency by damping (Appendix C). The FWHMs of both the red and blue modes converge to the same value ($\sim$ 145 GHz) at low temperatures. This reflects the finite Cooper plasmon relaxation rate $\gamma_{sc}$, which we attribute to both the ohmic contact and sidewall scattering \cite{kim2025}. Below $T_c$, but at frequencies above 2$\Delta(T)$, plasmons are composed of highly dissipative Bogoliubov quasiparticles. The temperature dependence of the superconducting gap causes the blue mode to cross from being Cooper-like at low temperatures to being Bogoliubov-like when the temperature is \SIrange{8.8}{11.1}{K}. The plasmon's resonance frequency is not sensitive to this transition, but the FWHM is, and it increases dramatically. The blue mode broadens with increasing temperature and becomes too broad and small to be fit properly at temperatures above \SI{9.9}{K}.

The red and the blues modes, together, take finite momentum, finite frequency snapshots of both the effective (fermionic) mass renormalization ($\Re[\Sigma_F]$) and (bosonic) plasmon lifetime ($\Im[\Sigma_B]$). The roughly constant resonance frequency implies that the metal to superconductor phase transition doesn't dramatically alter the quantum properties of the constituent electrons, but the increase of scattering rate when the plasmon crosses $2\Delta(T)$ shows that the plasmon lifetime drops when its character becomes Bogoliubov-like. This highlights another key point: the extraction of $\Re[\Sigma_F]$ and $\Im[\Sigma_B]$ of the red mode are independent from those of the blue mode. Having multiple, independent plasmons allows us to compare the self-energy's behavior at different momenta and frequencies, allowing us to isolate thermal or quantum changes to the superconductor itself.

\section{Discussion}\label{sec3} 

\begin{figure}[h]
\centering
    \includegraphics[width=0.5\textwidth]{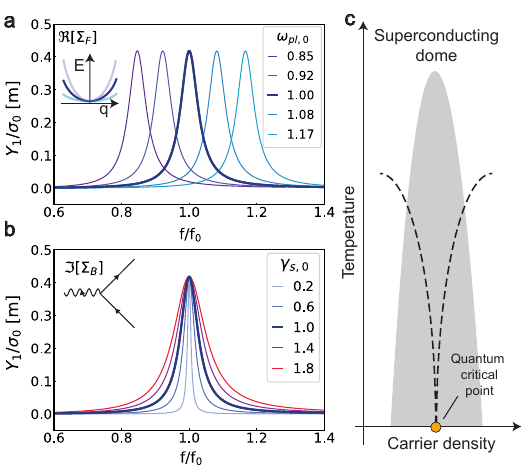}
    \caption{ \textbf{Independent resolution of the real and imaginary parts of the Cooper plasmon's self-energy} (a) Real part of the (fermionic) self-energy manifests as a shift in plasma frequency, which can come from carrier density (tunable by a gate in 2D systems) or renormalization of the effective mass by band structure changes or electronic correlations. Inset: band mass renormalization. (b) The imaginary part of the (bosonic) self-energy renormalizes the scattering rate, which manifests as a change to the FWHM. Inset: a sample lifetime renormalization by scattering. Both panels a and b were evaluated at a temperature of 0.1$T_c$. (c) An illustrative superconducting dome, with a pictured quantum critical point. }\label{Fig5_outro}
\end{figure}

The frequency and momentum at which Cooper plasmons probe the $\Sigma$ can be tuned by altering the cavity and transmission line geometry, allowing the targeting of specific superconducting features. For example, structural or interaction-driven renormalization of the  electron's effective mass would alter $\Re[\Sigma_F]$ and shift the Cooper plasmon's resonance frequency (Fig. \ref{Fig5_outro}a) at temperatures far below $T_c$. The supercarrier density controls $\Re[\Sigma_F]$, and that density can be gate-tuned in van der Waals systems and stoichiometrically tuned in other systems. Similarly, $\Im[\Sigma_B]$ controls the rate at which the Cooper plasmon scatters (Fig. \ref{Fig5_outro}b), thereby encoding scattering channels into other quasiparticle species and / or collective modes prominent at the quantum critical points seen at or near optimal doping in superconducting domes (such as cuprate superconductors) (Fig. \ref{Fig5_outro}c). In addition to equilibrium phase transitions, the direct extraction of $\Re[\Sigma_F]$ and $\Im[\Sigma_B]$ could also prove indispensable to the field of light-induced phase transitions \cite{Basov2017, bloch2022, Torre2021}, where the transient self-energy could contain singular or anomalous components. The robustness of our theory allows us to intentionally position one plasmon at a featureful frequency / momentum and one at an otherwise featureless frequency / momentum. One plasmon can serve as a reference for the other, ensuring that broadening and frequency shifts come from features in $\Sigma$ and not from trivial effects in the circuitry itself. 

Probing $\Sigma$ at finite momentum has many benefits beyond the phase mode. Many other superconducting collective modes, such as the Higgs, Bardasis-Schrieffer oscillations and others also lie outside the light cone and / or are infrared inactive at zero momentum \cite{shimano2020, Sun2020}, and could be targeted with microcavity resonances. In addition, the Eliashburg effect and light-matter coupling are enhanced by cavities \cite{curtis2019}, suggesting that superconducting self-cavities could be used to explore cavity modifications to superconductivity \cite{Schlawin2022, keren2025}.  

The near-field visibility of Cooper plasmons via the phase mode when the sample thickness descends below $\lambda_L$ implies that the superconductor circuit designs need to be conscious of the phase mode, particularly at high frequencies. Cooper plasmons could manifest as a large loss in superconducting circuits, such as transmon qubits, circuit QED devices, superconducting sensors, and kinetic inductance detectors \cite{blais2021, krantz2019}. Superconducting circuits rely on low-loss superconducting transmission lines, resonators, Josephson junctions, and other circuit elements embedded in microwave transmission lines. As the industry pushes toward and beyond 6G frequencies, which offer both high data transmission rates and less 1/f noise, it will become crucial to consider device geometries that suppress the absorption by Cooper plasmons. This can be done by using the analytic formulation presented here to keep the Cooper plasmonic resonances at frequencies far above a circuit's operating bandwidth. Though one could fabricate circuit elements with superconductors thicker than the London penetration depth, the 2D Cooper plasmons can become surface plasmons in 3D (see Appendix B). These surface plasmons will still be detected by the near-field nature of transmission line circuitry, and thus still parasitically attenuate propagating signals. It is imperative, therefore, that high frequency superconducting circuit engineers ensure that they avoid Cooper plasmons.

\section{Methods}\label{methods}  

\noindent \textbf{On-chip time domain THz spectroscopy:} On-chip time domain THz spectroscopy is an emerging technique used to directly extract the real and imaginary parts of the near-field THz conductivity. The low temperature (LT) GaAs emitter and `A' Auston switches are on a \SI{50}{\micro\meter}-thick cyclo-olefin polymer switch board - which only contains Auston switches and THz circuitry. The switch board used here is a model TeraLineX TD-800-CPS-50/30/50-3MM, supplied by Protemics GmbH. All Auston switches are excited with 3 mW of 780 nm optical power with 76 fs Gaussian laser pulses (250 MHz repetition rate). Samples can be evaporated, sputtered, or dry-transferred on interchangeable \SI{50}{\micro\meter} thick quartz sample boards. Measurement of the `A' Auston switch records both the signal just before it enters the switch board (directivity $d(t)$) and following reflection from the device on the sample board ($r(t)$). The optical alignment, automated optical realignment procedures, and analysis formulae are described in \cite{Potts2023a}. The time domain data is smoothed using a moving Savitsky-Golay filter, which removes noise above 6.9 THz. The low-frequency cutoff of the tool can be extended from 200 GHz \cite{Potts2023a} to 80 GHz because the effects of multi-bounce reflections can be shown to be negligible, with an upper bound of only 1.8\% error. Additional systematic errors associated with directivity calibration and referencing to normal state plasmons can be shown to be approximately 2\% and 4\% respectively, suggesting that features larger than $\sim$5\% are not experimental artifacts. For details on systematic and calibration errors, see Appendix C.\\

\noindent \textbf{Extraction of the admittance:} The superconductor's admittance was extracted via:

\begin{eqnarray}
    Y_{\text{sc}}(\omega)Z_0 = \frac{-2}{ 1 - \frac{R_{nm}(\omega)}{R_{\text{sc}}(\omega)}\frac{D_{\text{sc}}(\omega)}{D_{nm}(\omega) } \frac{2+Y_{nm} Z_0}{Y_{nm} Z_0} }
    \label{eqn2}.
\end{eqnarray}

Eqn.~\ref{eqn2} assumes a frequency-independent normal metal admittance $Y_{nm}$. In the normal metal phase of NbN, $\sigma_{\text{in}}(\omega)$ is characterized by a Drude model with $\gamma_{nm} \approx$ \SI{5.5}{THz} \cite{Sim2017}. This implies that the real conductivity at frequencies less than \SI{800}{GHz} is constant and has the same value as the DC conductivity, which was measured via DC transport \cite{Potts2023a}. We note that $\sigma_{\text{mc}}$ is not necessarily frequency-independent at high temperatures. The high resistivity of the `dirty' NbN metal at high temperatures, however, ensures that the system is in the ohmic limit, fixing $\sigma_{\text{mc}}(\omega)\rightarrow\sigma_{\text{in}}(\omega)$. This highlights an important point: the coupling between the CPS and a sample is not necessarily rigidly fixed to the ohmic or capacitive limit in the entirety of a parameter space. As the ratio $\sigma_{\text{ox}}/\sigma_{\text{in}}$ changes with temperature, magnetic field, gate voltage, pressure etc., the sample can transition between different regimes and one must be cognizant of this in order to correctly relate $\sigma_{\text{in}}(\omega)$ and $\sigma_{\text{mc}}(\omega)$.\\

\noindent \textbf{Sample nanofabrication:} The 151 nm thick NbN film was deposited according to \cite{Potts2023a}: magnetron sputtering with 300W power, 1.8 mTorr pressure, and 45/3.8 sccm Ar/N2 gas flow, resulting in 0.8 \AA/s deposition. Characterization with DC transport shows a residual resistivity ratio of 0.94 from 300K to 15K and a superconducting $T_c$ of 11.1K (resistance of 90\% of normal state resistance). The NbN extended beyond the transmission line by 350 $\mu m$ on each side, which does not affect the zeroth order estimate of the resonance frequencies (see Appendix C). Sample boards are made from 50 $\mu$m thick quartz. Thin quartz is secured to silicon carrier chips by thin crystal bond for nanofabrication. This is done by placing a small ($\sim$ 1 mm diameter) amount of crystal bond on cleaned silicon and heating the silicon to 120$^\circ$C. The sample board is then gently laid on the `dot' of crystal bond and gradually heated to 220$^\circ$C over the course of 20 minutes. Once the crystal bond has sufficiently reflowed, heat is turned off and the sample is allowed to cool, resulting in a $\sim$ 75 $\mu$m-thick crystal bond layer. Once nanofabrication is completed, the 50 $\mu$m quartz can be released by soaking the structure in heated N-Methylpyrrolidone for a few hours.\\

\noindent \textbf{Switch board - sample board alignment:} The switch board was placed under an optical microscope. The sample board was placed, sample size face down, on the switch board at the gap between A- and B- switches. A thin wire, controlled by a micropositioner, was used to push the sample board into position, such that the switch board CPS near the `A'-switch and sample board CPS were aligned with no more than 10 $\mu$m error. A SEM metal clip was screwed into the board and used to compress the quartz sample board into the COP switch board. 
The effects of minor misalignment are quantitatively analyzed in Appendix F. \\

\noindent \textbf{Computational Electromagnetics:} Electrodynamic simulations were performed to confirm the nature of the ohmic contact (see Appendix H). All simulations were performed with numerical finite-difference time-domain (FDTD) techniques using CST studio. Auston switches were modeled by surface currents with the approximate time profile of experimentally-measured pulses \cite{michael2025, kusyak2025, Kipp2025b}.

\backmatter

\bmhead{Acknowledgements}

The authors thank B. Schulte, K. Kusyak, A. Cavalleri, G. Meier, Yifan Su, and R.D. Averitt for fruitful discussions. A.M.P., A.K.N., and A.F.Y recognize support for sample nanofabrication by the National Science Foundation (NSF) Materials Research Science and Engineering Center (MRSEC) at UC Santa Barbara, Award DMR 1720256, and by the Gordon and Betty Moore Foundation EPIQS program under award GBMF9471. Work at UCSB was supported by the Air Force Office of Scientific Research under award \#FA9550-24-1-0113 to A.F.Y. A portion of this work was performed in the UCSB Nanofabrication Facility, an open access laboratory. We acknowledge support from the Max Planck-New York Center for Non-Equilibrium Quantum Phenomena. G.K. acknowledges support by the German Research Foundation through the Cluster of Excellence CUI: Advanced Imaging of Matter (EXC 2056, project ID 390715994). M.W.D. and M.H.M. acknowledge support from the Alexander von Humboldt Foundation. H.M.B. acknowledges financial support from the European Union under the Marie Sklodowska-Curie grant agreement no. 101062921 (Twist-TOC). A portion of this work was performed at the Institute for Terahertz Science and Technology (ITST) at UCSB.

\section*{Declarations}

\bmhead{Supplementary information}

This article is accompanied by supplementary material which contains the details of the breakdown of the Anderson-Higgs mechanism with dimensionality and Coulomb screening, details of the transition between electrodynamically 3D and 2D samples, our analytic self-cavity theory including the feedback factor, our generalized Zimmermann model, analysis of Cooper plasmon's scattering rate and Landau damping, experimental justification of the low-frequency spectrometer calibration, theoretical and numerical simulations showing the effects of ohmic contact, and extended details of the numerical fitting used to produce Fig. \ref{Fig4_fits}.

\bmhead{Author contributions}

The project was conceived by AMP and MHM. Theoretical analysis was done by MHM and AMP with assistance from GK, SML, HMB, JS, MWD, and FS. Numerical simulations were performed by TM. AMP and MHM processed experimental data with assistance from GK and LC. AKN and AMP collected experimental data, under the guidance of AFY. JWM and XYZ
supervised the project. AMP wrote the manuscript with the input of all coauthors.

\bmhead{Competing Interests}

The authors declare no competing interests.

\bmhead{Data availability}
The data that support the plots within this paper will be made available upon reasonable request.

\noindent

\bigskip


\bibliography{CooperPlasmons_references}

\end{document}